# Resolving the local distortions of Ising-like moments in magnetoelectric Ho-doped langasite.

A. Yu.Tikhanovskii[1,a], V.Yu. Ivanov[1], A.M. Kuzmenko[1], A. Stunault[2], O. Fabelo[2], E. Ressouche[3], V. Simonet[4], R. Ballou[4], I.A. Kibalin[2], A. Pimenov[6], A.A. Mukhin[1], and E. Constable[6,b]

[1]*Prokhorov General Physics Institute of the Russian Academy of Sciences, 119991 Moscow, Russia*
[2]*Institut Laue-Langevin, 71 avenue des Martyrs, 38042 Grenoble Cedex 9*
[3]*Université Grenoble Alpes, CEA, IRIG, MEM, MDN, 38000 Grenoble, France*
[4] *Institut Néel, CNRS and Université Grenoble Alpes, F-38000 Grenoble, France*
[6]*Institute of Solid State Physics, TU Wien, 1040 Vienna, Austria*

The magnetic properties of Ho- doped langasites $(La:Ho)_3Ga_5SiO_{14}$ are dominated by the Ising-like magnetic moments of the $Ho^{3+}$ ions. In their saturated regime, the induced magnetic state breaks both time and space inversion symmetries, leading to a novel linear magnetoelectric effect. However, due to distortions induced by a shared Ga/Si occupancy of the *2d* sites, resolving the microscopic nature of the magnetic configuration remains a difficult task. Here we combine polarized neutron diffraction and angular dependent magnetization experiments to determine the local distortions of the $Ho^{3+}$ magnetic moments in doped langasites $(La_{1-x}Ho_x)_3Ga_5SiO_{14}$ with $x \approx 0.015$ and $x \approx 0.045$. We propose a model for a field-induced magnetic configuration with arbitrary orientations of the local Ising axis of $Ho^{3+}$ in distorted positions. The operations of broken local $C_2$ symmetry and rotations around the trigonal $C_3$ axis connect different sites, restoring the global $P321$ symmetry of the crystal and simplifying the description of the magnetic properties. The superposition of two distorted $Ho^{3+}$ positions connected by $C_2$ symmetry determines the local magnetic susceptibility tensor, which no longer appears Ising-like at low fields.

## I.     INTRODUCTION

Compounds with the langasite structure ($La_3Ga_5SiO_{14}$) [1–4] have been extensively studied for some years now and are primarily well-known for their strong piezoelectric and nonlinear optical properties [5–7]. The existence of orthogonal third and second order axes in the *P*321 space group leads to a number of unusual functional properties. For example, uniaxial pressure along the second-order axis breaks the trigonal symmetry and induces an electric polarization [6]. Significant interstitial spaces between $La^{3+}$ dodecahedra and $Ga^{3+}$ octahedra make them highly susceptible to stress [7] appreciably increasing the mechanical coupling, which is higher than that of quartz [8]. These properties, combined with structural stability over a wide range of temperatures, make langasite particularly attractive for use in piezoelectric sensors [9].

As shown in Fig. 1, the unit cell of langasites consists of two alternating layers. The first layer consists of $La^{3+}$ dodecahedra at the *3e* positions with $C_2$ symmetry and of $Ga^{3+}$ octahedra at the *1a* positions with $D_3$ symmetry. The second layer consists of $Ga^{3+}$ and $Si^{3+}/Ga^{3+}$ tetrahedra occupying the *3f* positions with $C_2$ symmetry and *2d* positions with $C_3$ symmetry, respectively. Oxygen ions ($O^{2-}$) are located at the vertices of these polyhedra. Magnetic ions in the langasite lattice can lead to non-trivial magnetic structures along with unconventional magnetoelectric properties. For example, the well-studied iron langasites (such as $Ba_3NbFe_3Si_2O_{14}$) order antiferromagnetically at $T_N \sim 27$ K into a triangular spiral magnetic structure with a double chirality [10, 11]. In external magnetic fields they exhibit exotic chiral-dependent magnetoelectric coupling [12–14]. However, the presence of such magnetic structures complicates the study of the microscopic mechanisms underlying the magnetic and magnetoelectric properties of langasites, especially in external magnetic fields. From this point of view, the rare-earth langasites $R_3Ga_5SiO_{14}$ (R = La, Pr, Nd, Tb, Dy, Ho,…) are of particular interest. For instance, the well-studied compounds $Nd_3Ga_5SiO_{14}$ and $Pr_3Ga_5SiO_{14}$ [15–20] remain paramagnetic down to very low temperatures (30 mK) within a distorted Kagome – like lattice [15, 16]. The crystal structure of R-langasites is stable only for the light rare-earth ions with R=La, Nd, Pr. Starting from Sm, a phase separation occurs with additional garnet-like phases [21].

---

[a] tikhanovskii@phystech.edu
[b] evan.constable@tuwien.ac.at

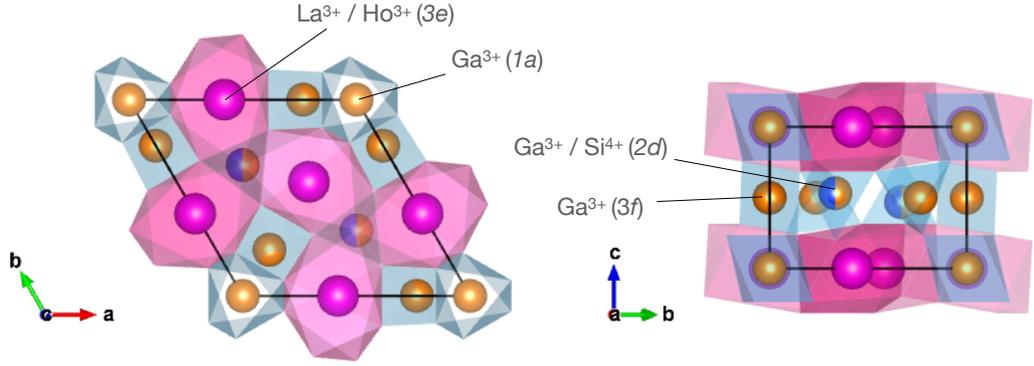

FIG. 1. Crystal structure of langasites. The crystallographic positions with ion labels and Wyckoff notation are indicated.

The compounds containing heavy rare-earth ions with large magnetic moments (Tb, Dy, Ho, ...) are therefore only stable at low concentrations of the rare-earth ions. In such compounds one can expect a strong perturbation of the ion's ground state within the crystalline field [22–28]. This determines the local magnetic anisotropy of the ions, along with macroscopic magnetic and other physical properties of the crystal. In addition, due to the non-centrosymmetric crystal structure of langasites, they may exhibit significant magnetoelectric coupling, as seen in rare-earth ferro- and alumoborates [22–27], which possess a similar trigonal structure ($R32$ or $P3_121$ space groups).

The magnetic and magnetoelectric properties of a heavy $Ho^{3+}$ doped langasite $(La_{1-x}Ho_x)_3Ga_5SiO_{14}$ (x≈0.015) (Ho-LGS) were investigated for the first time in Ref. [29]. The quasi-doublet ground state of the non-Kramers $Ho^{3+}$ ion determines the macroscopic magnetic properties of the crystal. It is largely separated from other excited levels, which makes it an Ising-like ground state. Different local orientations of the $Ho^{3+}$ Ising axes in various positions determine its magnetic anisotropy. This picture of the $Ho^{3+}$ ground state was proposed in Ref. [29]. However, it does not capture all relationships between the local and macroscopic magnetization, which may indeed be more complicated. This is most likely because there is a shared occupancy of the 2d positions featuring Ga and Si ions of different size (see Fig. 1). This breaks the local $C_2$ symmetry of rare-earth ions at the 3e positions and leads to a deviation of local magnetization axes (i.e., the Ising axes of $Ho^{3+}$) from the orientations allowed by the unperturbed $C_2$ symmetry.

Polarized neutron diffraction (without polarization analysis), involves measuring a neutron flipping ratio (i.e., the intensity ratio of neutrons scattered with incoming spins parallel and antiparallel to an external magnetic field [30]). It is an exceptionally powerful tool for studying magnetization processes at local crystallographic positions. Unlike conventional magnetization measurements, which provide information about the averaged macroscopic magnetization value, polarized neutron diffraction determines the magnetization density within the unit cell. This allows an investigation of magnetization processes occurring at the microscopic level. In particular, this method has been applied to study the local susceptibility anisotropy of rare-earth ions in pyrochlore titanates $R_2Ti_2O_7$ (R = Ho, Dy, Tb) with a frustrated spin ice magnetic structure [31–36].

In this work we will demonstrate that polarized neutron diffraction experiments are highly effective in a complementary study using both neutron scattering and magnetization in external magnetic fields, even for a very low fraction of magnetic ions within the material. To investigate the magnetization processes at the $Ho^{3+}$ local positions and the influence of local distortions on their behavior, we performed a comprehensive experimental and theoretical study of the magnetic properties of Ho substituted paramagnetic langasite. The local susceptibility was examined by polarized neutron diffraction in a Ho-LGS (x≈0.045) single crystal. Magnetization measurements were carried out under a magnetic field applied along the principal crystallographic directions. We also explored the details of the magnetization anisotropy under magnetic fields rotated within the *ab\**, *ac*, and *b\*c* planes. To explain the experimental data, we propose a model that considers the influence of local distortions on the ground state of the Ising-like $Ho^{3+}$ ion, with local positions connected by $C_2$ and $C_3$ symmetry operations, which preserves the global symmetry of the crystal. This model allows us to quantitatively describe the unique magnetic anisotropy of Ho-LGS. In particular, we determine the parameters of the local distortions (deviations and distributions of the Ising axis) and establish their correlation with the macroscopic magnetic properties of the crystal. These refinements are crucial to the microscopic analysis and consistent with a quantitative description of the observed Ho-LGS magnetoelectric properties (i.e. various field-induced components of electric polarization), which are currently under investigation and will be presented in further publication.

## II. METHODS

Doped langasite crystals were grown by two different methods: the Ho-LGS crystal with x≈0.015 was grown using the Czochralski method by B.V. Mill [1,2]. The crystal with x≈0.045 was grown using the floating-zone method by A.M. Balbashov. We determined the quality of the crystals by X-ray analysis and by scanning electron microscopy in the z-contrast mode. In both samples, only the langasite phase was detected. The magnetic properties of the langasites were studied using a MPMS-50 (Quantum Design) magnetometer in magnetic fields up to 5 T and at temperatures from 1.9 K to 60 K.

The neutron diffraction studies were performed on the x ≈ 0.045 Ho-LGS sample at the Institute Laue-Langevin (ILL), Grenoble, France. The nuclear structure factor and extinction parameters were first characterised at 2 K using the D9 diffractometer operated at λ = 0.836 Å, produced using the (220) plane of a Cu crystal monochromator, the λ/2 contamination was avoided by using an Er absorption filter in the transmission geometry. Integrated intensities of structural Bragg peaks were collected with standard *ω*-scans for low-*q* reflections and *ω*-2*θ*-scans for medium and high-*q* reflections. Polarised neutron diffraction was then performed on the D3 instrument with λ = 0.825 Å. Measurements were undertaken at 5 K in a magnetic field of 0.5 T, i.e. in the linear regime. Flipping ratios were measured on a pre-aligned crystal, with directions *b*, *b\** and *c*. For this, the OrientExpress Laue instrument at ILL was used [37]. In total 933 reflections were measured, with the external magnetic field aligned parallel to the *b* (305 reflections), *b\** (484 reflections) and *c* (144 reflections) axes.

Analytically, the experimentally obtained flipping ratio, *R*, is expressed as,
$$R = \frac{I^+}{I^-} \quad (1)$$
where $I^+$ and $I^-$ are the scattering intensities of neutrons polarized parallel and anti-parallel to the external magnetic field respectively.

The relationship between the flipping ratio and the local susceptibility parameters is laid out in detail by Gukasov and Brown in Ref. [30]. A brief summary is also provided in the supplementary material. Ultimately, the local susceptibility tensor is obtained by a least-squares refinement of the experimental flipping ratios with respect to Eq. 1.

## III. EXPERIMENT

### A. Magnetic properties

Angular dependence of the magnetization was measured for various temperatures and in different magnetic fields rotating in the $ab^*$, $b^*c$ and $ac$ planes. In addition, we measured the field dependence of the magnetization along different crystallographic directions in the temperature range from 1.9 K to 90 K.

The observed angular dependences provide the most relevant information. At low temperatures and in strong magnetic fields, where the magnetic moments along local axes are almost (quasi-) saturated (the Zeeman quasi-doublet splitting exceeds the thermal energy), we observe noticeable anisotropy of the magnetization (Fig. 2 a-c) associated with the peculiarities of the local Ising axis orientation of $Ho^{3+}$ ions.

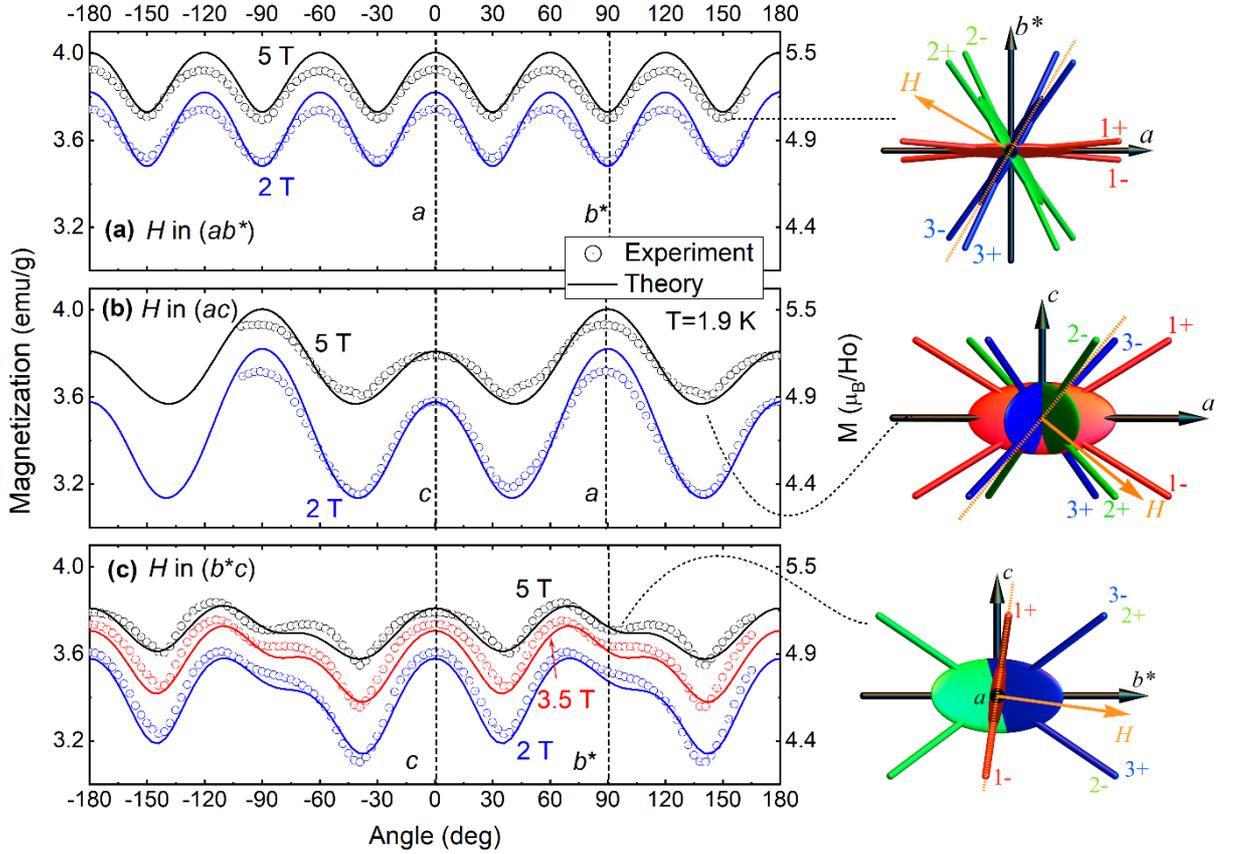

FIG. 2. Angular dependence of the magnetization in Ho-LGS ($x \approx 0.045$) in (a) $ab^*$, (b) $ac$, and (c) $b^*c$ planes at $T=1.9$ K and in external field of $\mu_0 H = 5$T – black, 3.5 T – red, and 2 T – blue. Open symbols represent experimental data, solid lines are theoretical predictions detailed in the text. The pictograms in the right column show the orientation of the magnetic field at the points corresponding to the minima in the angular dependences, i.e. where the field is perpendicular to some of the Ising axes of $Ho^{3+}$ ions. $1^\pm$, $2^\pm$, and $3^\pm$ represent the orientations of six equivalent magnetic axes of $Ho^{3+}$, respectively.

In the quasi-saturated regime ($\mu_0 H > 1$ T and $T=1.9$ K), a 60-degree anisotropy was observed in the $ab^*$ (hexagonal) plane, with a maximum along the $a$-axis and a minimum along the $b^*$-axis (i. e. perpendicular to the second-order $a$-axis) (see Fig. 2a). In the $ac$ plane an asymmetric minimum occurs between the two non-equivalent maxima for $H||a$ and $H||c$ (Fig. 2b). In the $b^*c$ plane we observe the most characteristic benchmarks of the magnetization anisotropy. When the field is oriented in this plane, the angular dependence is asymmetric with respect to the local maximum along the $c$-axis (Fig. 2c). For the clockwise and anticlockwise rotation of the magnetic field relative to the *c*-axis, the neighboring minima have different depths, and the following maxima have different amplitudes. The *b**-axis is located near one of the local minima but does not coincide with it.

The depth of the magnetization minima decreases with increasing temperature and the angular dependencies become smoother, which is also observed for fields below 1 T. In the high-temperature and low-field regime, the magnetization in the *ac* and *b*c* planes follows a $M \sim \sin 2\theta_H$ dependence, while there is no anisotropy in the *ab** plane. The characteristic features in the angular dependence of the magnetization are the same for both compounds with x≈0.045 and x≈0.015 (see Supplementary Material).

In agreement with the data above, the low temperature magnetization curves demonstrate an anisotropic behavior. The magnetization saturates along local axes around a field of ~1 T. The Van-Vleck contribution and the deviation from Ising behavior of the magnetic moments that tend to bend toward the magnetic field (see the Theory Section) leads to a small effective susceptibility for magnetic fields above 1 T and along the principal crystallographic directions *a*, *b**, and *c* (Fig. 3a). In contrast, the magnetization curves with the field aligned along one of the minima in the angular dependence (i.e. when the field is orthogonal to one of the most probable Ising axes) display a stronger additional contribution to the magnetization in the saturation regime (see Fig. 3b – *Hc*-39°*b*). This additional contribution is characterized by a steeper slope and by a reduction of angular modulation with increasing magnetic field strength (see Fig. 2(c) for fields 2 T, 3.5 T, and 5 T). Finally, Fig. 6, shown later in the Theory Section, demonstrates how the magnetization asymmetry in the *b*c* plane is compatible with the crystallographic symmetry *P*321 and how it is associated with a deviation of the local magnetic ellipsoids from the *ac* plane.

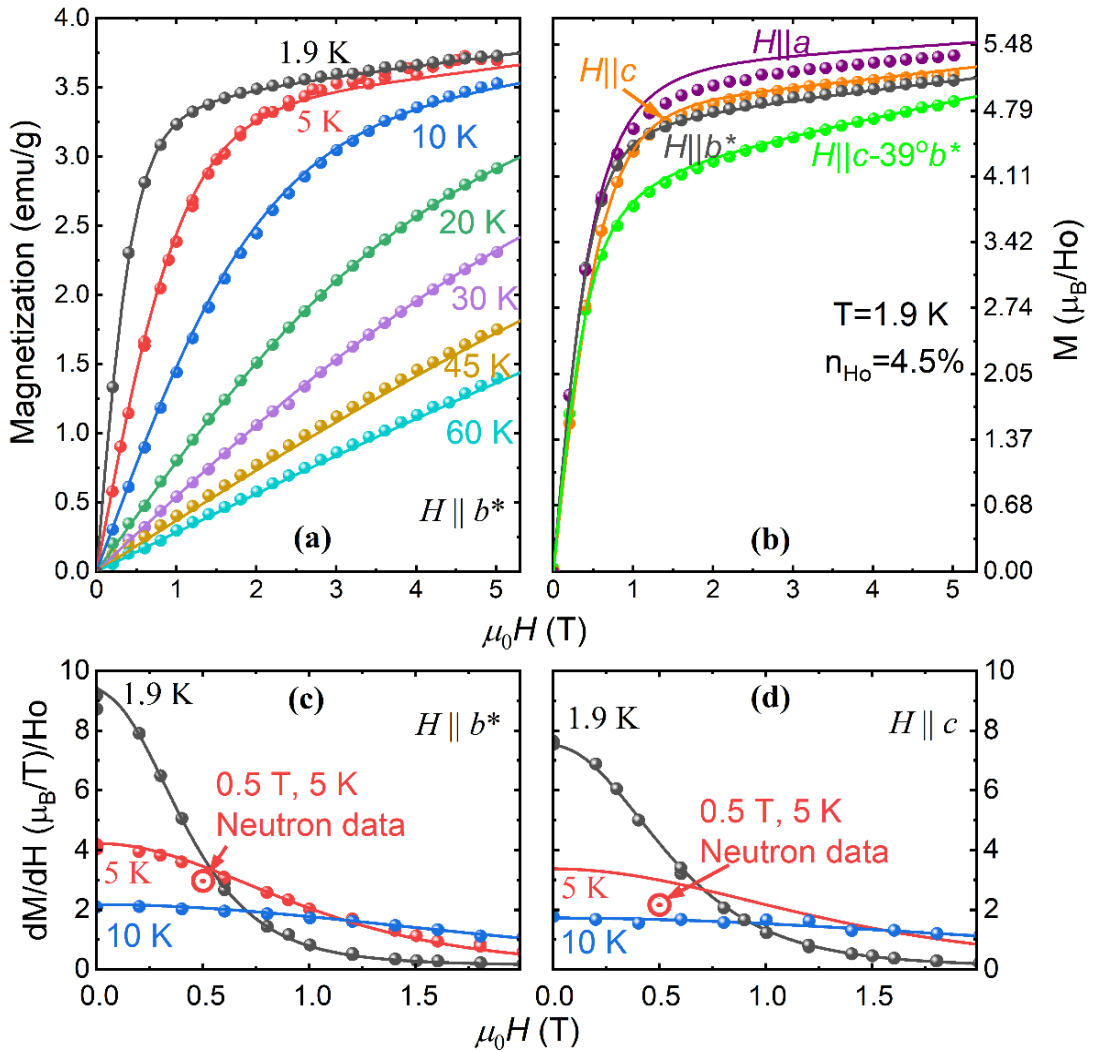

FIG. 3. (a) Magnetization curves of Ho-LGS (x≈0.045) for $H||b^*$ and at temperatures between 1.9 K and 60 K. (b) Comparison of the magnetization curves at $T=1.9$ K for $H \parallel a, b^*, c$, and $c$-39°$b^*$ directions. The last direction corresponds to the minimum of the angular dependence in the $b^*c$ plane (see Fig. 2(c)). (c) Comparison of the susceptibility observed from magnetization and from neutron data for $H \parallel b^*$ and (d) for $H \parallel c$. Symbols represent experimental data, solid lines are theoretical calculations (see sections 4.1-4.3).

## B. Neutron Scattering

Flipping ratios for a magnetic field of $\mu_0 H = 0.5$ T applied along the $b$, $b^*$ and $c$ directions were refined simultaneously using the ChiLSQ subroutine of the Cambridge Crystallography Subroutine Library (CCSL). A second refinement was also performed using the Crystallographic Python Library (CrysPy), with both libraries giving equivalent outputs. Later, these refinements were further validated using the Mag2Pol software. The magnetic form factor was approximated at the dipole limit as $f(k) = \langle J_0(k)\rangle + (\frac{2}{g_L} - 1)\langle J_2(k)\rangle$, where $\langle J_{0,2}(k)\rangle$ are radial integrals of the $Ho^{3+}$ electronic wavefunction and $g_L = 5/4$ is the Landé g-factor, accounting for the orbital contribution to the magnetic moment (see supplementary material for details [38]). The refinement quality for the three different applied magnetic field directions is shown in Fig. 4.

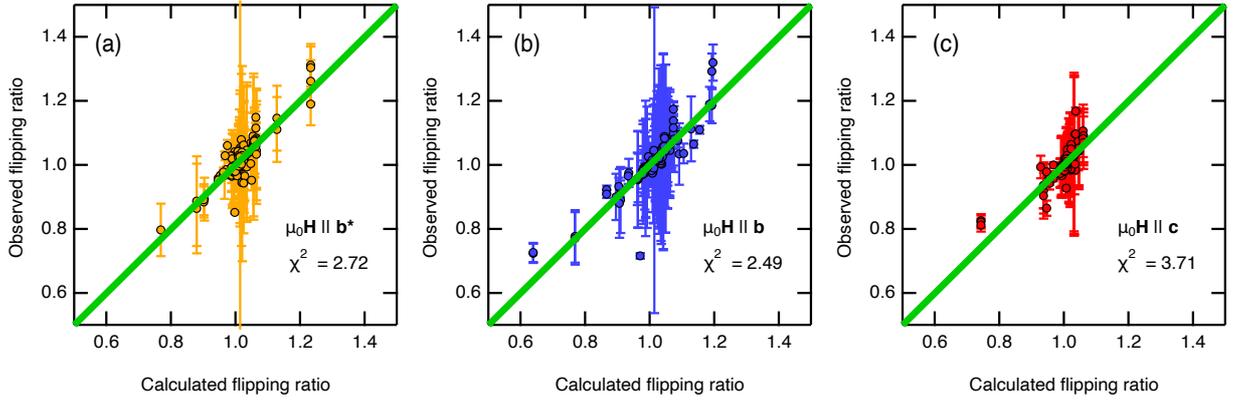

FIG. 4. Neutron flipping ratio refinement using CCSL and dipole approximation for applied fields of 0.5 T along the (a) $b^*$, (b) $b$, and (c) $c$ crystallographic directions. The measurements were performed at a sample temperature of 5 K.

The refined susceptibility tensor for a single $Ho^{3+}$ ion rounded to two decimal places was found to be,

$$\hat{\chi} = \begin{pmatrix} 5.403(1) & 0.00(5) & 0.002(1) \\ \chi_{12} & 0.2(2) & 0.5(4) \\ \chi_{13} & \chi_{23} & 2.4(1) \end{pmatrix} \mu_B/T, \quad (2)$$

within the orthogonal coordinate system $\{x\}||\mathbf{a}$, $\{y\}||\mathbf{b}^*$, and $\{z\}||\mathbf{c}$. Diagonalizing the tensor gives the following eigenvectors $\mathbf{X} = \mathbf{a}$, $\mathbf{Y} = 0.98\mathbf{b}^* - 0.21\mathbf{c}$, and $\mathbf{Z} = 0.21\mathbf{b}^* + 0.98\mathbf{c}$, such that,

$$\chi_x \mathbf{X} + \chi_y \mathbf{Y} + \chi_z \mathbf{Z} = 5.40\mathbf{X} + 0.12\mathbf{Y} + 2.50\mathbf{Z} \; [\mu_B/T]. \quad (3)$$

A graphical representation of the corresponding magnetization ellipsoid is shown in Fig. 5 (a,c). It consists of an ellipse, mostly in the $ac$ plane, with a slight rotation of 13° ± 7° around the $a$-axis towards the $b^*$-axis.

The bulk susceptibility per Ho ion is then determined by,

$$\chi_{ij} = \frac{1}{N_{\text{Ho}}} \sum_{k=1}^{k=3} \sum_{l=1}^{l=3} R_{ik} R_{jl} \chi_{kl}, \quad (4)$$

where $N_{\text{Ho}}$ is the number of Ho ions in the unit cell and $1 < i \leq 3$ and $1 < j \leq 3$ for all $\tilde{R}$ operators in the $C_3$ point group. Here we apply the inverse property of symmetry rotations $R_{ij} = R_{ji}^{-1}$. The resulting bulk susceptibility per Ho ion is therefore

$$\hat{\chi}_{\text{Bulk}} = \frac{1}{3} \begin{pmatrix} 8.45 & 0 & 0 \\ 0 & 8.45 & 0 \\ 0 & 0 & 7.17 \end{pmatrix} [\mu_B/\text{Ho}]/T. \quad (5)$$

The result is in fair agreement with the measured magnetic field dependence in the linear regime as shown in Fig. 3.

We note that the ellipsoid of the local magnetic susceptibility extracted from the neutron data does not correspond to the expected Ising-like behavior of Ho$^{3+}$ ions proposed in Ref. [29]. Therefore, in the proceeding Theory Section we develop a more sophisticated model. It starts from the Ising-like character of Ho$^{3+}$ ions and accounts for the observed susceptibility ellipsoid assuming a distribution of locally-distorted positions.

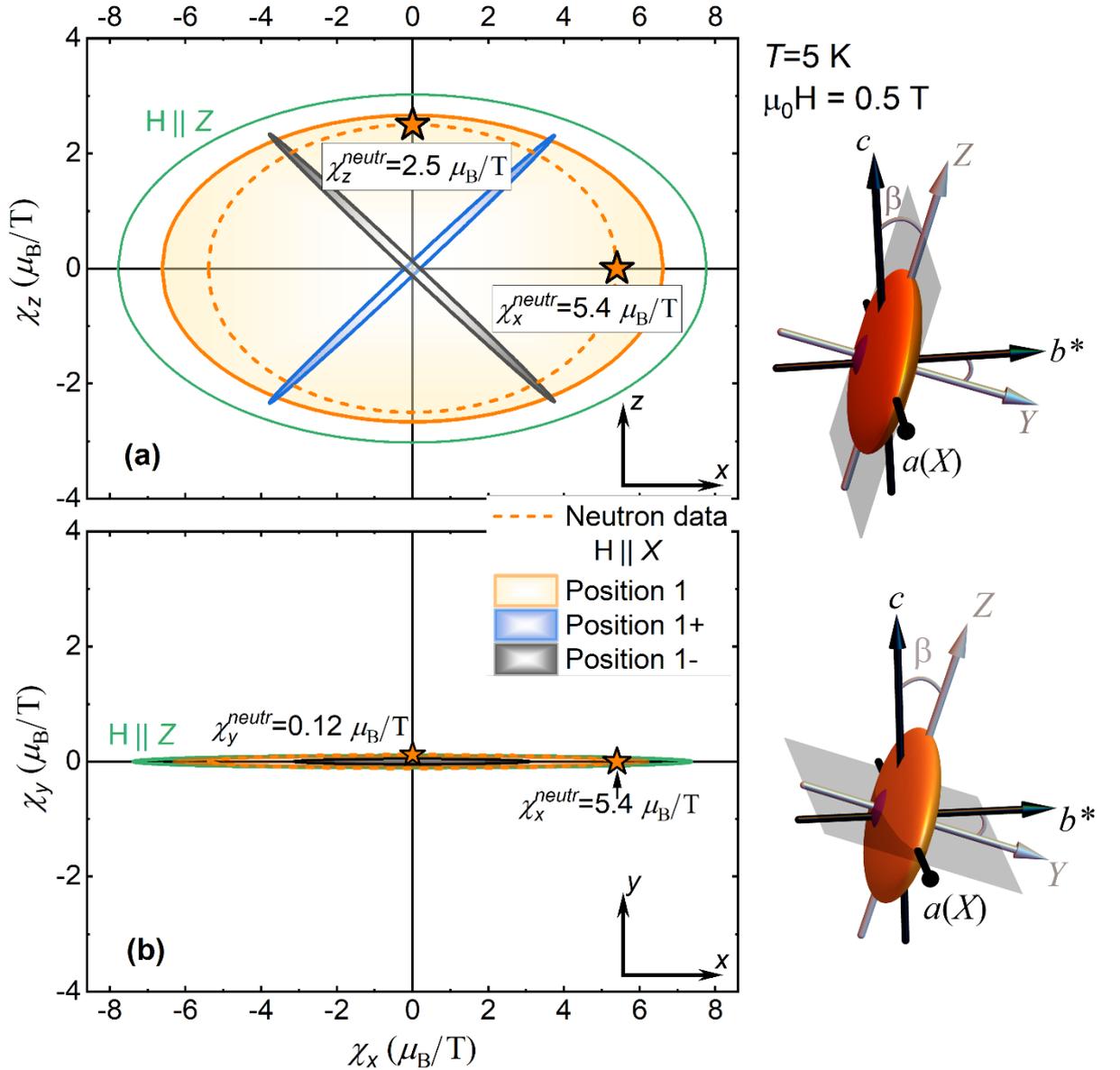

FIG. 5. Calculated and measured magnetic susceptibility ellipsoids of $Ho^{3+}$ in Ho-LGS (x≈0.045) represented in the local coordinate system of a pair of distorted crystallographic positions connected by the $C_2$ symmetry. Projections of the ellipsoids on the planes of the local coordinate system are shown for $XZ$ (a) and $XY$ (b). The external magnetic field is $\mu_0H = 0.5$ T and the temperature is $T = 5$ K. The right-most column shows the schematic orientations of the susceptibility ellipsoids within the crystallographic and local coordinates. The projection planes $XZ$ (a) and $XY$ (b) are indicated by the gray planes. The crystallographic coordinate system is rotated with respect to the local $XYZ$ axes by an angle $\bar{\beta} \approx 4°\pm 2°$ around the second-order axis ($C_2 \| a$). Narrow blue and black ellipses show the Ising-like susceptibility of two most probable "+" and "-" directions coupled by the $C_2$ symmetry along the $a$-axis. The orange and green solid lines represent the model ellipses of susceptibility accounting for small non-linear field-contributions for static fields $H \| X$ (orange) and $H \| Z$ (green) for both distorted positions (see details in 4.4 section). Dotted lines represent the projections of the magnetic ellipsoid obtained from polarized neutron diffraction. Stars denote the susceptibility values along the local $XYZ$ axes refined from neutron measurements.

## IV.  THEORY AND DISCUSSION

### A.  Model of disorder-induced anisotropy distribution

In Ho-LGS, the $Ho^{3+}$ magnetic ions occupy three low-symmetry positions with a $C_2$ symmetry axis aligned to one of the three crystallographic directions ($a$, $b$, $-a-b$). They remain in a paramagnetic state down to low temperatures. However, the shared occupation of Ga and Si at the 2d positions leads to a breaking of the local $C_2$ symmetry in the rare-earth environment. The possible presence of a superlattice in LGS [39, 40] suggests the existence of some energetically more favorable states (i.e., arrangements of Ga and Si). It is impossible to unambiguously identify them from the present data only, but we assume that rigorous microscopic calculation may determine the set of allowed configurations.

A crystal electric field of low symmetry ($C_1$) splits the ground multiplet $^5I_8$ into $2J+1 = 17$ singlets. The two lowest neighboring energy levels (a quasi-doublet with a small splitting of $\Delta_{cf} \approx 3$ K) primarily determine the magnetic properties of the $Ho^{3+}$ non-Kramers ion in langasite (see also Ref. [29]). Such a ground state results in an Ising-like behavior. The matrix element of the magnetic moment operator between the wave functions of this quasi-doublet determines the magnetization of the $Ho^{3+}$ ions, which exhibit strong anisotropy (see, for example, Ref. [41]). This is consistent with the low temperature behavior of the magnetization curves, showing a large slope in low fields, and a fast saturation at $\mu_0H \sim 1$ T (Fig. 3).

To describe the magnetic properties of Ho-LGS, we used an approach similar to Ref. [29] based on a spin Hamiltonian that includes splitting of the quasi-doublet in the crystal electric field, along with a Zeeman energy and Van-Vleck contributions:

$$H_{eff}^{(i)} = \Delta_{cf}\sigma_z^{(i)} + \boldsymbol{m}_i\boldsymbol{H}\sigma_y^{(i)} - \frac{1}{2}\boldsymbol{H}\hat{\chi}_{VV}^{(i)}\boldsymbol{H}. \quad (6)$$

Here $\sigma_{z,y}^i$ are the Pauli matrices at the $i$-th position, $\boldsymbol{m}_i\boldsymbol{H}=\mu_0\boldsymbol{n}_i\boldsymbol{H}=-i\mu_B g_J<A|\hat{\boldsymbol{J}}_i\boldsymbol{H}|B>$ (where $|A>$ and $|B>$ denote the wave functions of the quasi-doublet), $\boldsymbol{n}_i$ is a unit vector of an arbitrarily oriented Ising axis, $\boldsymbol{H}$ is the external magnetic field and $\hat{\chi}_{VV}^{(i)}$ is the Van-Vleck susceptibility matrix. The energy levels of the quasi-doublet at the $i$-th position, obtained by diagonalization of the spin Hamiltonian, are equal to $\pm\varepsilon_i$, where $\varepsilon_i = \sqrt{(\boldsymbol{m}_i\boldsymbol{H})^2 + \Delta_{cf}^2}$, and determine the magnetization of $Ho^{3+}$ ions (see below).

Next, we consider possible distributions of local magnetization directions (Ising axes) and their relationship to different positions within the crystal. Let's assume that the magnetization direction (Ising axis) is determined by an arbitrary unit vector $\boldsymbol{n}(\alpha,\beta)=(\cos\alpha, \sin\alpha\sin\beta, \sin\alpha\cos\beta)$. Here $\alpha$ is its deviation from the $a$-axis, and $\beta$ is the angle between the projection of the Ising axis in the $b^*c$ plane and the $c$-axis (as in Fig. 6). Due to the random distribution of Ga/Si ions breaking the local $C_2$ symmetry of the Ising axis, the angles $\alpha$ and $\beta$, should also be randomly distributed around their most probable values $\bar{\alpha}$ and $\bar{\beta}$, which can be determined by modeling of the magnetic properties. In this work we use a two-dimensional Gaussian distribution $\rho(\alpha,\beta) = Exp[-\frac{(\alpha-\overline{\alpha})^2}{2\delta_\alpha^2}]Exp[-\frac{(\beta-\overline{\beta})^2}{2\delta_\beta^2}]/2\pi\delta_\alpha\delta_\beta$ for the Ising axes directions with dispersions $\delta_\alpha$ and $\delta_\beta$.

To preserve a global $P321$ symmetry of the Ho-LGS crystal we assume the presence of another distorted position coupled with the first one by the $C_2$ symmetry operation. Taking into account that the angle $\overline{\alpha}$ is invariant while $\overline{\beta}$ transforms to $\overline{\beta}-\pi$ under the $C_2$ symmetry operation, the distribution function then becomes $\rho'(\alpha,\beta) = Exp[-\frac{(\alpha-\overline{\alpha})^2}{2\delta_\alpha^2}]Exp[-\frac{(\beta-(\overline{\beta}-\pi))^2}{2\delta_\beta^2}]/2\pi\delta_\alpha\delta_\beta$. Thus, the superposition of $\rho$ and $\rho'$ in distorted positions determines the distribution function $\rho_+=(\rho+\rho')/2$ of the averaged local positions possessing the global $C_2$ symmetry.

The orientation of the Ising axes and the distribution functions in other positions are determined by ±120° rotation around the trigonal ($C_3$) axis. Let us now denote the position with an arbitrarily oriented Ising axis $\mathbf{n}_1(\alpha,\beta) = (\cos\alpha, \sin\alpha\sin\beta, \sin\alpha\cos\beta)$ and consider the distribution function $\rho_+$ at position "1". Here, the form of $\rho_+$ reflects the presence of the two most probable directions (coupled by $C_2$ symmetry). The additional positions "2" and "3" are related to "1" by an extra $C_3$ symmetry operation. The rotation matrices $\widehat{3}_c^\pm$ determine the pair of Ising axis directions $\mathbf{n}_{2,3}(\alpha,\beta) = \widehat{3}_c^\pm \mathbf{n}_1(\alpha,\beta)$ in positions "2" and "3". The distribution function $\rho_+$ remains unchanged under the symmetry operations. Thus, we introduce a three-sublattice model with the distribution function $\rho_+$. It preserves the global symmetry $P321$ of the crystal despite its local deviations.

Since all three sublattices, "1", "2" and "3" include two components, they can be also treated as pairwise split into six sublattices, $1^\pm$, $2^\pm$, $3^\pm$. The most probable Ising directions for $1^\pm$ are $\mathbf{n}_{1+}(\overline{\alpha},\overline{\beta})$ and $\mathbf{n}_{1-}(\overline{\alpha},\overline{\beta}-\pi)$, with $\mathbf{n}_{2+,3+}(\overline{\alpha},\overline{\beta}) = \widehat{3}_c^\pm \mathbf{n}_{1+}(\overline{\alpha},\overline{\beta})$, $\mathbf{n}_{2-,3-}(\overline{\alpha},\overline{\beta}) = \widehat{3}_c^\pm \mathbf{n}_{1-}(\overline{\alpha},\overline{\beta}-\pi)$ for $2^\pm$ and $3^\pm$, respectively. They are connected by the set of threefold symmetry axis rotations on $\mathbf{n}_{1\pm}$ (Fig. 6). The distribution function $\rho_+$ is the same in all positions "1", "2" and "3". For the interpretation of magnetization curves, we will employ the six-sublattice approach, while, for neutron data, it is more convenient to use the three-sublattice model with a generalized distribution function. Note that if we consider the undistorted $Ho^{3+}$ positions with local $C_2$ symmetry, it is not possible to describe the angular and field dependences of the magnetization consistently (see Ref. [29]).

In the frame of this model, we can now define the total free energy of the two level ($\pm\varepsilon_i$) system and calculate the average magnetization considering three types of magnetic positions with the distribution $\rho_+$ of Ising axes:

$$\mathbf{M} = \frac{1}{3}n_{Ho}\sum_{i=1,2,3}\left(\iint [\mathbf{m}_i(\mathbf{H}\mathbf{m}_i) + 2(\Delta_{cf} + \mathbf{H}\hat{p}_i\mathbf{H})\hat{p}_i\mathbf{H}]\frac{\tanh(\varepsilon_i/k_B T)}{\varepsilon_i}\rho_+ d\alpha d\beta + \hat{\chi}_{VV}^{(i)}\mathbf{H}\right). \qquad (7)$$

Here, $\mathbf{m}_i=\mu_0\mathbf{n}_i$, $\hat{p}_i$, $\varepsilon_i$ and $\rho_+$ depend on $\alpha$, $\beta$, $n_{Ho}$ is the $Ho^{3+}$ ion concentration, and $k_B$ is the Boltzmann constant. We also take into account a weak renormalization of the crystal field splitting by magnetic field $\Delta_{cf} \rightarrow \Delta_{cf} + \mathbf{H}\hat{p}_i\mathbf{H}$ due to an influence of excited $Ho^{3+}$ states (see also Ref. [42]). It is characterized by the local magnetic tensor $\hat{p}'$, which can be represented in the crystallographic coordinate system by the rotation matrices $\hat{p}_i = \hat{C}_i^{-1}\hat{p}'\hat{C}_i$ [38]. The $\hat{p}'$ tensor depends on the matrix elements of the magnetic moment between the ground and excited $Ho^{3+}$ crystal field states and leads to a nonzero susceptibility when the magnetic field and Ising axis are orthogonal.

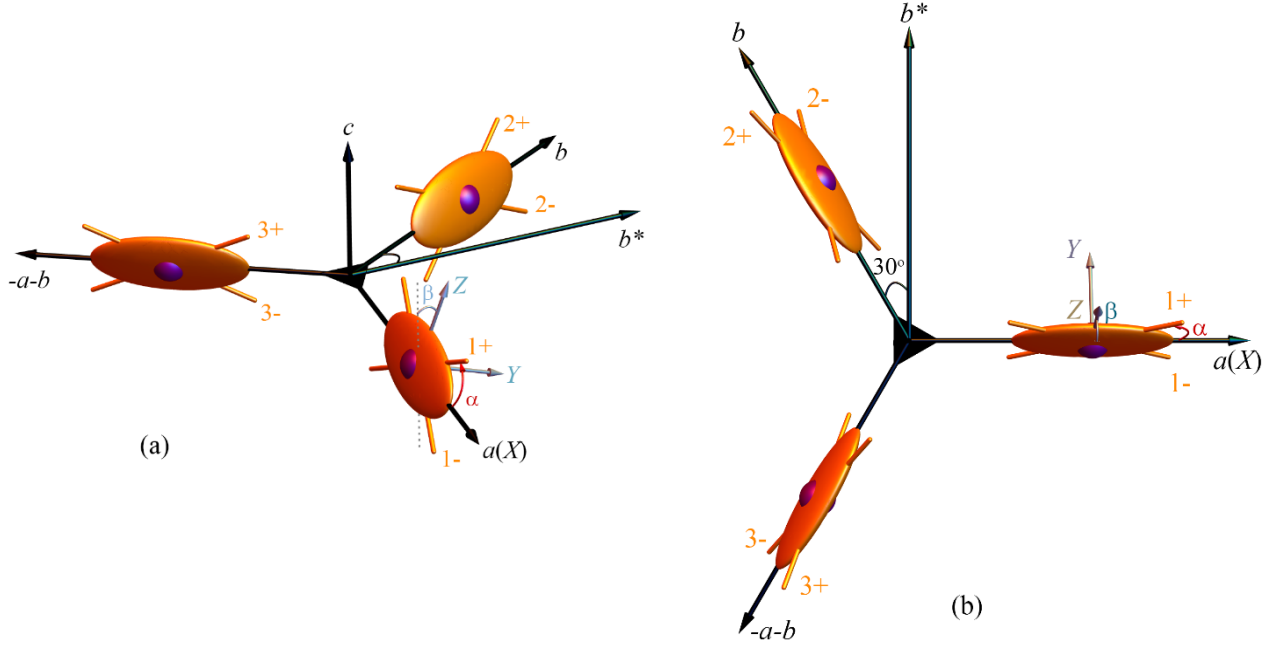

FIG. 6. Schematic representation of susceptibility ellipsoids and the most probable directions of the Ising axes. (a) - in a three-dimensional view and (b) - projected onto the *ab* plane. The numbers ±1, ±2, ±3 denote the most probable directions of the Ising axes in the corresponding positions. The local coordinate system *XYZ*, rotated by an angle *β* around the second-order axis, is shown for position "1" only. The angle *α* represents the deviation of the most probable directions of the Ising axes from the *a*-axis.

### B. Angular dependence of magnetization

The model of Ising axis directions qualitatively explains the presence of extrema in the angular dependence of magnetization. The distinctive features of the Ising-like behavior of $Ho^{3+}$ and of the disorder-induced distribution of their anisotropy axis are the most prominent in the quasi-saturated regime ($\mu_0 H > 1$T and $T < 5$K). Specifically, when the magnetic field is oriented perpendicular to one of the Ising axes, minima appear in the angular dependence of magnetization (Fig. 2a-c).

For magnetic fields rotated in the *ab\** plane, the magnetization behaves as $M \sim \sin 6\varphi$, and the equivalence of magnetic moment orientations parallel and antiparallel to the Ising axes leads to the presence of a 60-degree periodicity. We note that when the magnetic field is rotated by 90 degrees in the *ab\** plane, three minima should be observed, for example, in the range from 0 to +90°, when the magnetic field $H$ is orthogonal to the most probable axes $n_{2+}$, $n_{2-}$, and $n_{1-}$ (Fig. 2 right). However, the directions $n_{2+}$ and $n_{2-}$ are close to each other, thus making the corresponding minima indistinguishable (Fig. 2a).

When the field rotates by 180 degrees in the *ac* plane, two minima with slight asymmetry are observed (Fig. 2b). These minima are the combinations of three weaker minima. For example, when the magnetic field rotates from the *c*-axis to the *a*-axis, it becomes orthogonal to the axes $n_{1-}$, $n_{2+}$ and $n_{3+}$ at the angles $\theta_H \approx 32°$, 49° and 53°, respectively (right column in Fig. 2). These features are not resolved in the experiment as they are smoothed out due to the random distribution. In the angular dependences, the maximum for $H\|c$ arises due to the deviation of the Ising axes from the *ab\** plane. From the magnetization ratio for $H\|c$ and $H\|a$ we estimate the deviation angle as $\bar{\alpha} \approx 32°$ (Fig. 6a).

Further characteristic features related to the orientation of the Ising axes are observed for magnetic fields in the *b\*c* plane. In this plane, the angles at which the field is orthogonal to the axes $n_{3-}$ and $n_{2+}$, $n_{1+}$ and $n_{1-}$, $n_{2-}$ and $n_{3+}$ coincide. As a result, only three distinct minima are observed in the 180-degree angular dependencies (Fig. 2c). The local minimum near $H\|b^*$ occurs when the field is orthogonal to $\mathbf{n}_{1+}$ and $\mathbf{n}_{1-}$ ($\theta_H \approx 94°$, right column in Fig. 2). The inequality $\theta_H \neq 90°$ and the asymmetry of the angular dependence in the $b^*c$ plane, when the sign of the magnetic field projection onto the $c$-axis changes ($\theta_H = 90°\pm\delta$, Fig. 2c), indicates a deviation of the local $XZ$ plane (which includes $\mathbf{n}_{1+}$ and $\mathbf{n}_{1-}$, see Fig. 6b) from the $ac$ plane. The estimation of this deviation is $\overline{\beta} \approx 4°\pm 2°$, which falls within the uncertainty bands of the angle determined by neutron refinements.

### C. Field induced magnetic moment distribution

Using the magnetization equation in terms of the magnetic structure model, the field dependences of magnetization were simulated as well (Fig. 5). From these data we can estimate the concentration of the magnetic ions in both samples ($n_{Ho}$ = 1.53%, 4.45%) and the magnetic moment of the doublet ($\mu_0$=9.4$\mu_B$) for $H \| a, b^*$ and $c$. The magnetic properties for the sample with $n_{Ho}$ = 1.53% can be described self-consistently without additional parameters [38].

As discussed above, for $H \| a, b^*$ and $c$, the magnetization slope in the saturation regime and for the sample with n=4.45% is rather small. In contrast, when the magnetic field is orthogonal to one of the Ising axes (Fig. 5b, $H \| c$-39°$b^*$), this slope increases significantly. We believe that this is related to the deviation of Ho$^{3+}$ ions from pure Ising behavior. The $\hat{p}'$ components that determine this slope are equal to $p_x$=0, $p_y$= 17·10$^{-2}$ $\mu_B$/T, and $p_z$=42.1·10$^{-2}$ $\mu_B$/T. A finite slope in the magnetization along the $a$, $b^*$ and $c$ - axes in the saturation regime indicates the presence of Van-Vleck contributions to the susceptibility. Summed over three positions it can be estimated as

$$\chi_{VVa} = \chi_{VVb^*} = \frac{1}{3} n_{Ho} \left|\sum_{i=1,2,3} \hat{\chi}_{VV}^{(i)} \mathbf{H}/H\right|_{a,b^*} = 2.7\cdot 10^{-2} \ \mu_B/T, \quad \text{and}$$

$$\chi_{VVc} = \frac{1}{3} n_{Ho} \left|\sum_{i=1,2,3} \hat{\chi}_{VV}^{(i)} \mathbf{H}/H\right|_c = 1.2\cdot 10^{-2} \ \mu_B/T.$$

The distribution of the Ising axes leads to a broadening of the minima in the angular dependencies of magnetization. In combination with deviations from the Ising behavior, a quantitative description of the depth of minima in Figs. 2 a-c is obtained for the dispersion $\delta_\alpha$= 6.7° and $\delta_\beta$= 8°.

### D. Local magnetic susceptibility

The magnetic susceptibility calculated using the parameters obtained from the model above is also in agreement with the susceptibility obtained from the polarized neutron diffraction measurements (Fig. 5). We calculated the magnetic susceptibilities of the local positions, which are superpositions of susceptibilities from two distorted Ho$^{3+}$ positions connected by $C_2$ symmetry. We used the definition of magnetization $M$ from Eq. (7) to determine the effective susceptibility $\hat{\chi}$ of position "1". The increment of the induced magnetization by the infinitesimal change of the magnetic field $d\mathbf{h}$ is equal to $d\mathbf{M}^{(1)} = \mathbf{M}^{(1)}(\mathbf{H}+d\mathbf{h}) - \mathbf{M}^{(1)}(\mathbf{H})$.

For the low field regime $\mu_0 H \ll k_B T, \Delta_{cf}$ one can linearize $d\mathbf{M}^{(1)}$ by the increment of static field $d\mathbf{h}$ and introduce it's from as:

$$d\mathbf{M}^{(1)} = n_{Ho}\hat{\chi}^{(1)}(\mathbf{H})d\mathbf{h}. \qquad (8)$$

Here the effective local susceptibility is,

$$\chi_{\xi\eta}^{(1)}(\mathbf{H}) = \iint \left(m_\xi m_\eta - \frac{1}{4}\left(\frac{mH}{k_B T}\right)^2 (2m_\xi m_\eta + m_\xi^2)\right)(k_B T)^{-1} \rho_+ d\alpha d\beta + \chi_{VV}^{\xi\eta} \qquad (9)$$

with $\xi, \eta = a, b^*, c$. For simplicity, we omitted renormalization of the crystal field splitting by magnetic field due to the influence of excited Ho$^{3+}$ states.

The linearization with respect to *H* perfectly works in the low field and high temperature regime. In the crystallographic coordinate system the susceptibility tensor for T = 5 K is equal to

$$\hat{\chi}^{(1)} = \iint \begin{pmatrix} m_x^2 & 0 & 0 \\ 0 & m_y^2 & m_y m_z \\ 0 & m_y m_z & m_z^2 \end{pmatrix} (k_\mathrm{B}T)^{-1} \rho \, d\alpha d\beta + \hat{\chi}_{VV}^{(1)} =$$

$$= \begin{pmatrix} 8.47 & 0 & 0 \\ 0 & 0.08 & 0.23 \\ 0 & 0.23 & 3.32 \end{pmatrix} (\mu_B/\mathrm{Ho})/T \qquad (10)$$

where we took the $C_2$ symmetry into account, which cancels certain components and allows us to reduce the $\rho_+$ distribution function for two sites to just one site, $\rho$. In this case, the susceptibility in position "1" is described by the components $\hat{\chi}^{(1)}$ (Fig. 5), and the susceptibilities of "2" and "3" are connected to "1" by ±120° rotation (Fig. 6). The rotation around the *a*-axis by an angle $\beta^{rot} = \frac{1}{2}\arctan(\int \sin 2\beta \, \rho d\beta / \int \cos 2\beta \, \rho d\beta) \approx 4°$ diagonalizes the susceptibility matrix, which is equal to $\bar{\beta}$ and, taking uncertainties into account, is in agreement with $\beta_{\mathrm{neutr}} \approx 13°\pm 7°$. Thus, in the local coordinate system, obtained by rotating the crystallographic coordinates system by an angle $\beta^{rot}$ around *X* parallel to the second order axis $C_2$ (*a*) (as shown in the right column of Fig. 5), the susceptibility is well described by an ellipse, which is a cross-section of the ellipsoid $x^2/\chi_{xx}^2 + y^2/\chi_{yy}^2 + z^2/\chi_{zz}^2 = 1$. The calculated eigenvector describing the magnetic ellipse of a single Ho site is given by

$$\chi_{xx}X + \chi_{yy}Y + \chi_{zz}Z = 8.47X + 0.07Y + 3.34Z \; [\mu_B/\mathrm{Ho}]/T, \qquad (11)$$

which is larger than that determined by the neutron refinements (3).

The reason for this is likely due to nonlinearities already present at 0.5 T. Specifically there is a decrease in the susceptibility under increasing applied fields (Fig 3 c,d). For magnetic fields $(\boldsymbol{mH})^2 \sim (k_\mathrm{B}T)^2$ the additional quadric term given by the static magnetic field *H* must also be considered. It reduces the value of $\hat{\chi}^{(1)}$ according to Eq. (9) and makes positions "1", "2" and "3" nonequivalent. Using Eq. (9) we calculated the eigenvector describing the magnetic ellipse of a single Ho site in the nonlinear regime (T=5 K and $\mu_0 H$=0.5 T) for *H* || *X* and *Z* respectively obtaining,

$$\chi_{xx}X + \chi_{yy}Y + \chi_{zz}Z = 7.79X + 0.06Y + 3.02Z \; [\mu_B/\mathrm{Ho}]/T$$

$$\chi_{xx}X + \chi_{yy}Y + \chi_{zz}Z = 6.62X + 0.05Y + 2.67Z \; [\mu_B/\mathrm{Ho}]/T. \qquad (12)$$

The values are now closer to those determined by the neutron refinement, but differ for *H* || *X* and *Z* (Fig 5a). Here the $\beta^{rot}$ dependence on *H* and *T* is negligible. We highlight that the neutron refinements were performed simultaneously on data collected for *H* || *b**, *b* and *c* and considering all three Ho positions coupled by $C_3$ symmetry. However, due the non-linear terms mentioned above, the response will depend on the field direction with respect to the local axes of the ellipsoids in different positions. . Thus, in this particular case, the refined susceptibility tensor obtained from the neutron data ($\hat{\chi}^{neutr}$) should be interpreted as an averaged and effective susceptibility.

The non-zero susceptibility for $d\boldsymbol{h}$ || *Y* (Fig. 5b) is the result of the distribution of the Ising axes and deviations from the Ising behavior, with the second case not being taken into account in Eqs. (8)-(12). Accounting for these deviations doubles the value $\chi_{yy} \approx 0.10 \; (\mu_B/\mathrm{Ho})/T$. The corresponding ellipse in the *XY* plane calculated numerically is shown on Fig. 5b. There is a good agreement between the calculated and refined susceptibilities $\chi_{yy}$. However, we note that the effective susceptibility $\chi_{yy}^{neutr}$ has a large relative error because of its small value. Moreover, the Gaussian distribution of Ising axes is a model approximation, which can introduce errors in the calculated susceptibilities for fields orthogonal to the Ising axes. Finally, we did not account for the specifics of the wave functions of $\mathrm{Ho}^{3+}$ to estimate the components of $\hat{p}'$ which describe the deviation from the Ising behavior and strongly affect the $\chi_{yy}$ magnitude. A rigorous mathematical treatment considering various configurations of Ga and Si ions and calculating the spectrum of $\mathrm{Ho}^{3+}$ ions would enable a more accurate modeling of the magnetic properties. However, in this

work we used a simpler approach and were still able to obtain a good agreement between model and experiment.

In summary, the magnetization experiments provide the following approximate distribution of the magnetic anisotropy of Ho-LGS. It consists of two Ising axes $n_{1+}$ and $n_{1-}$, at distorted positions connected by $C_2$ symmetry and rotated from the local $a$-axis by the angles $\bar{\alpha} \approx \pm 32°$, respectively. The plane built of $n_{1+}$ and $n_{1-}$ roughly coincide with the ac-plane but is rotated away from it by the angle $\bar{\beta} \approx 4°$ (see Fig. 6). Both angles $\alpha$ and $\beta$ reveal Gaussian distributions with dispersions $\delta_\alpha = 6.7°$ and $\delta_\beta = 8°$, respectively. Four other $Ho^{3+}$ positions are obtained from $n_{1+}$ and $n_{1-}$ applying the $C_3$ symmetry operation along the $c$-axis.

Taking into account the experimental errors and the model assumptions, the susceptibility agrees rather well with the refinement obtained from polarized neutron diffraction measurements. A graphical comparison of the susceptibility ellipsoids from magnetization and from neutron experiments is shown in Fig. 5.

An important aspect obtained from the magnetization data is the observable influence of the local directions and the distributions of the Ising axes. To obtain further details of this magnetic configuration, more sophisticated experiments and rigorous theoretical calculations are necessary.

## V. CONCLUSION

Using a comprehensive experimental and theoretical approach, we studied the effect of $Ho^{3+}$ local magnetic orientations and their distortions on the macroscopic magnetic properties of doped langasites $(La_{1-x}Ho_x)_3Ga_5SiO_{14}$ with $x \approx 0.015$ and $x \approx 0.045$.

The local magnetic susceptibility at local $Ho^{3+}$ positions was determined by flipping ratio measurements using polarized neutron diffraction. Taking the $C_2$ symmetry of the local $3e$ $Ho^{3+}$ positions into account, the magnetic ellipsoids of susceptibility were determined by refinement of the neutron data. The highest susceptibility (the major axis of the ellipsoid) occurs for fields directed along the $C_2$ axes $X$ ($H \parallel a, b, -a-b$). The next-largest susceptibility is seen in a field directed along the local $Z$ axis, that deviates from the $ac$-plane plane by an angle $\beta_{neutr} \approx 13° \pm 7°$. The local susceptibility along the $Y$-axis, which is perpendicular to the $X$ and $Z$ axes, is an order of magnitude smaller than the other two. Here the refinement of the polarized neutron data assumed a local $C_2$ symmetry and provided a good approach to capture the magnetic density at local crystallographic positions. The insight into the details of the rare-earth sites with violated $C_2$ symmetry was obtained from the analysis of the angular dependent magnetization in the $ab*$, $b*c$, and $ac$ planes especially in the quasi-saturated regime ($\mu_0 H > 1$ T).

We propose a simple model of the field-induced magnetic structure assuming a quasi-doublet ground state of the non-Kramers $Ho^{3+}$ ions in a crystal field environment. This includes a perturbation of the Ising-like behavior with an arbitrary orientation of the local Ising axis in distorted positions due to shared occupation of the neighboring Ga/Si 2d sites in langasites. We argue that an averaged distribution of the Ising axes for the distorted $Ho^{3+}$ positions are coupled by local $C_2$ symmetry and by the rotations around the trigonal $c$-axis, which restores the global $P321$ symmetry of the crystal. Within these assumptions it is possible to quantitatively describe the observed global magnetic properties and the ellipsoids of the local susceptibility. As a result, microscopic characteristics of the $Ho^{3+}$ ground state, such as the most probable local orientations, the amplitude and distribution of the magnetic moments could be determined. These characteristics do not depend on the Ho-doping value of the studied langasite crystals, thus confirming the validity of the approach.

Finally, similar influence of local magnetic distortions on the macroscopic properties in doped langasite crystals reveals characteristic features that can be studied in other systems as well. More detailed insight into the distribution of local magnetic properties requires rigorous microscopic calculations and may also necessitate the development of new experimental approaches.

## ACKNOWLEDGMENTS

We thank Navid Quershi for validating our neutron refinement results using the Mag2Pol software. This work was supported by the Russian Science Foundation (Project No. 22-42-05004) and the Austrian Science Fund (Projects No. P32404 and No. I5539).

# SUPPLEMENTAL MATERIAL

## Rotation of the local magnetic tensor

The local $a$-axis in position "1" is the second order-axis. In the spin coordinate system, the local $x$-axis is parallel to the Ising axis. The connection between the local magnetic and the crystallographic coordinate systems can be represented as:

$$(x', y', z') = \hat{B}(\beta)\hat{A}(\alpha)(a, b^*, c). \qquad (S1)$$

Here, $\hat{A}(\alpha) = \begin{pmatrix} \cos\alpha & 0 & \sin\alpha \\ 0 & 1 & 0 \\ -\sin\alpha & 0 & \cos\alpha \end{pmatrix}$ is the rotation around $b^*$ by the angle $\alpha$ and $\hat{B}(\beta) = \begin{pmatrix} 1 & 0 & 0 \\ 0 & \cos\beta & -\sin\beta \\ 0 & \sin\beta & \cos\beta \end{pmatrix}$ is the second rotation around $x'$ by the angle $\beta$. Therefore, the local coordinate system of the $Ho^{3+}$ ion is characterized by $\alpha$ and $\beta$.

Taking this into account the local susceptibility tensor $\hat{p}'$ that describes the deviation from the Ising behavior, is equal to

$$\hat{p}_i = \hat{C}_i^{-1}\hat{p}'\hat{C}_i, \qquad (S2)$$

in the crystallographic coordinate system. Here $i$ is the position number, $\hat{C}_1 = \hat{B}(\beta)\hat{A}(\alpha)$, $\hat{C}_{2/3} = \hat{B}(\beta)\hat{A}(\alpha)\hat{3}_c^{\pm}$, $\hat{3}_c^{\pm}$ represents an additional rotation by $\pm 120°$ around the trigonal axis. Further arguments in favor of this magnetic configuration can be drawn from the magneto-electric properties of Ho-LGS, which will be analyzed in the subsequent work.

## Magnetization data for Ho-langasite with x≈0.015

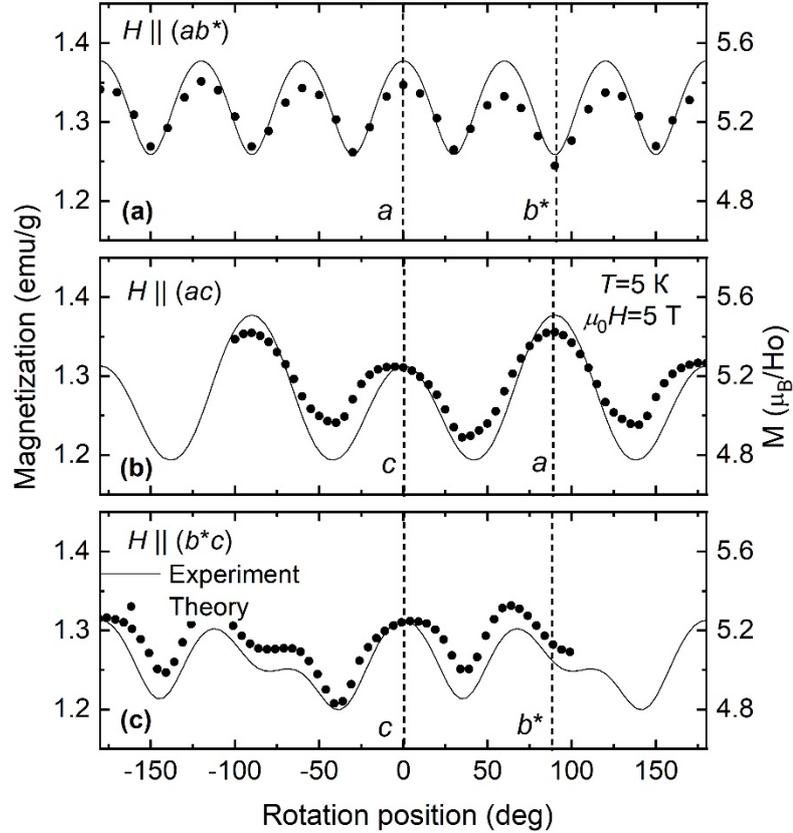

Supplementary FIG. 1. Angular dependencies of the magnetization of Ho-LGS ($x \approx 0.015$) in the (a) $ab^*$, (b) $ac$, and (c) $b^*c$ planes at $T = 5$ K and for $\mu_0 H = 5$ T. Symbols represent the experimental data, solid lines correspond to the theory.

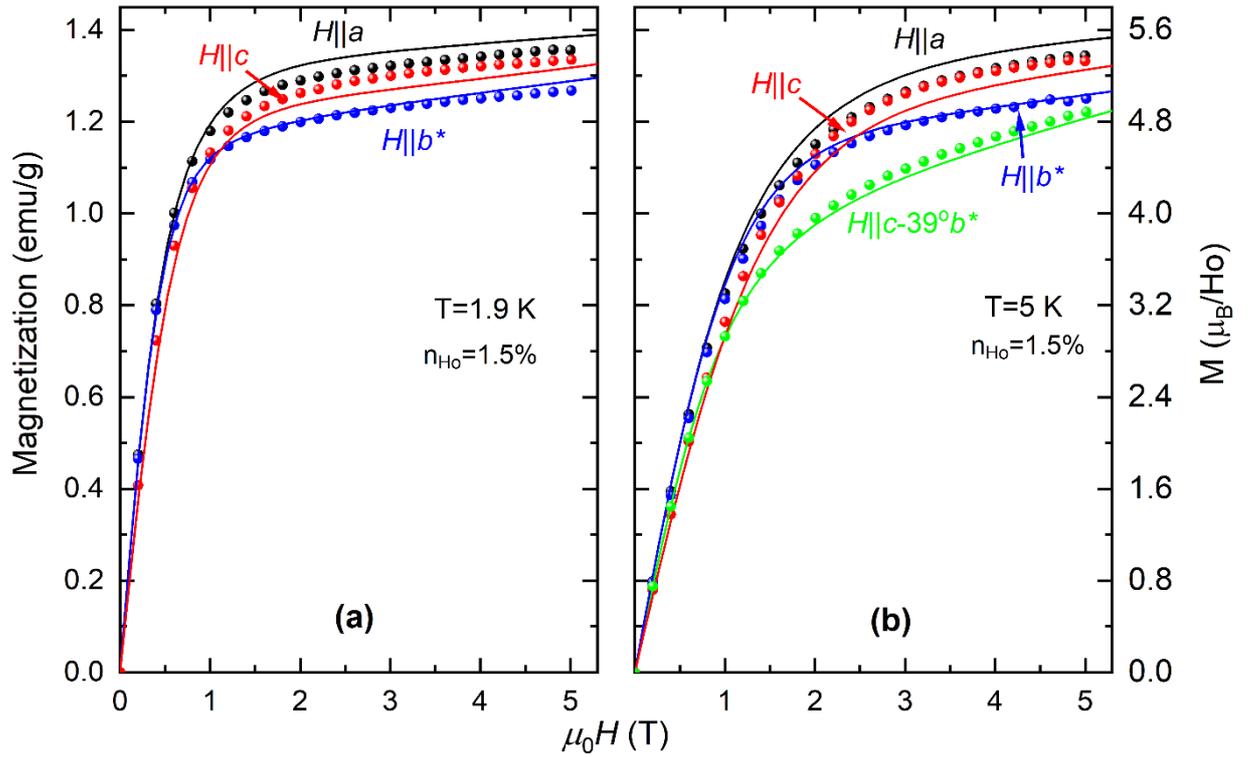

Supplementary FIG. 2. Magnetization curves of Ho-LGS (x≈0.015) for (a) $H \parallel a, b^*, c$, and $T = 1.9$ K, and (b) $H \parallel a, b^*, c, c\text{-}39°b^*$ and T = 5 K. The direction $c\text{-}39°b^*$ corresponds to the minimum of the angular dependence in the $b^*c$ plane. Symbols represent experimental data, solid lines correspond to the theory.

# Relationship between neutron flipping ratio and magnetic susceptibility

This section summaries the work of Gukasov and Brown in Ref. [30]. The scattering intensity for polarized neutrons is given by,

$$I = N^2 + 2\mathbf{P_0} \cdot (N'\mathbf{M'_\perp} + N''\mathbf{M''_\perp}) + \mathbf{M}_\perp^2. \quad (S5)$$

Here, $N$ is the nuclear structure factor with real ($N'$) and imaginary ($N''$) components, while $\mathbf{M'_\perp}$ and $\mathbf{M''_\perp}$ are the real and imaginary components of the magnetic interaction vector $\mathbf{M_\perp}$ defined as $\mathbf{M_\perp}(\mathbf{k}) = \mathbf{k} \times \mathbf{M}(\mathbf{k}) \times \mathbf{k}$. $\mathbf{P_0}$ is the incident neutron beam polarization vector. The experimentally measured quantity is the flipping ratio,

$$R = \frac{I^+}{I^-} \quad (S6)$$

where $I^+$ and $I^-$ are the scattering intensity of neutrons polarized parallel and anti-parallel to an external magnetic field $H$.

In order to relate the flipping ratio to the magnetic susceptibility tensor, we consider a single magnetic ion ($a$) with an anisotropic susceptibility resulting from its local environment. For low magnetic fields in the linear regime, the magnetization induced by an external field $H$ is given by

$$\mathbf{M}^a = \chi^a \mathbf{H}.$$

The local magnetization of a symmetrically equivalent atom (b) is then given by,

$$\mathbf{M}^b = \chi^b \mathbf{H} = \widetilde{R}\chi^a \widetilde{R}^{-1} \mathbf{H}, \quad (S7)$$

where $\{\widetilde{R}:t\}$ is a symmetry operator.

Taking into account the relative positions of the magnetic ions within the unit cell, and assuming a spherical distribution of the magnetic moment around each atom, the magnetization distribution in the unit cell is given by,

$$\mathbf{M}(\mathbf{r}) = \sum_p \widetilde{R}_p \chi^a \widetilde{R}_p^{-1} \mathbf{H} \, \rho(\mathbf{r} - \widetilde{R}_p \mathbf{r}_a - \mathbf{t}_p). \quad (S8)$$

Here the sum is over all $N_g$ operators $\{\widetilde{R}_p:t_p\}$ in the space group. In our analysis we consider a magnetic form factor with a dipolar approximation of the form:

$$f(k) = \langle J_0(k)\rangle + \left(\frac{2}{g_L} - 1\right)\langle J_2(k)\rangle, \quad (S9)$$

with,

$$\langle j_\kappa(k)\rangle = \int_0^\infty r^2 R_{nl}^2(r) j_\kappa(kr)\, dr. \quad (S10)$$

Here, $R_{nl} = \langle r|nl\rangle$ are the radial parts of the electronic wavefunction and $j_\kappa(kr)$ are the $\kappa$-order Bessel functions. The second term in S9 accounts for the orbital contribution to the magnetic moment. Here, $g_L$ is the Landé g-factor calculated for the ground multiplet quantum number at low temperature. In the case of $Ho^{3+}$ (featuring 10 $f$-electrons) the ground multiplet is $^5I_8$ (S = 2, L= 6, J = 8) giving a value $g_L$= 5/4. The values for $\langle J_\kappa(k)\rangle$ have been computed for trivalent ions such as $Ho^{3+}$ from solutions of Dirac-Fock Hamitonians and are tabulated in the International Table of Crystallography [43].

The corresponding magnetic structure factor is then given by,

$$\mathbf{M}(\mathbf{k}) = \frac{1}{N_a} f(k) \sum_p \widetilde{R}\chi^a \widetilde{R}^{-1} \mathbf{H} e^{i\mathbf{k}\cdot(\widetilde{R}_p \mathbf{r}_a + \mathbf{t}_p)} \quad (S11)$$

Here $N_a$ is the number of operators $q$ for which $\mathbf{R}_p \mathbf{r}_a + \mathbf{t}_p = \mathbf{r}_a$, such that is the multiplicity of the site $a$ and its symmetry is determined by the point group $Q$, formed by the rotational parts of the set of operators $q$. With this, the flipping ratio $R$ can be expressed in terms of the magnetic susceptibility tensor allowing a least squares refinement of the experimental measurements.